\documentclass[eqsecnum,aps,superscriptaddress,11pt ]{revtex4}

\begin{document}

\title{Accurate relativistic {\it ab initio} study of interstellar forbidden lines of singly ionized Zinc using the coupled cluster approach}

\author{Gopal Dixit$^1$, B. K. Sahoo$^2$,  R. K. Chaudhuri$^3$, and Sonjoy Majumder$^1$ \\
\vspace*{0.3cm}
{\it $^1$Department of Physics, Indian Institute of Technology-Madras, Chennai-600 036, India \\
$^2$ Max Planck Institute for the Physics of Complex Systems, D-01187 Dresden, 
Germany\\
$^3$ Indian Institute of Astrophysics, Bangalore-34, India}
}


\begin{abstract}
\noindent
In this work, the {\it ab initio} calculations have been carried out to study the 
transition probabilities corresponding to the forbidden transitions of astrophysically 
important electromagnetic transitions in the singly ionized zinc, Zn II. Many important 
electron correlations are considered to all-orders using the relativistic coupled-cluster theory. 
Calculated ionization potentials  are compared with the experimental values, 
wherever available. To our knowledge, oscillator strengths of the magnetic dipole and  electric quardupole transitions 
are estimated for the first time. The transitions 
span in the range of ultraviolet, visible, and near infrared regions and are crucial for the
astrophysical observations. 
\end{abstract}
\maketitle

\section{Introduction}
With the advancement of satellite based observations, we can get the information about the abundance of elements
in the interstellar space which provides a probe of the nature and composition of the interstellar medium. The data
obtained with the Goddard High Resolution Spectrograph (GHRS), the Hubble Space Telescope (HST) and other satellites
based on the spectroscopic facilities exhibit very high signal-to-noise ratios and allow routine measurements of very weak
lines (e.g. forbidden lines) with attainable accuracy. 

Generally allowed electric dipole transitions (E1) are important because of their strengths and easily accesible for the
incoming observations. However, forbidden transitions, like electric quadrupole (E2) and magnetic dipole transitions (M1), have
also great importance in the astrophysics even though they are weak, because they carry valuable informations like thermal doppler effects of heavenly bodies. 
The forbidden lines provide important clues in other areas of astrophysics, beacuse of the long lifetime of the upper 
state against radiative decay. These lines are particularly sensitive to the collisional de-excitation and serve as indicators
of electron density and temparature, $N_e$ and $T_e$, in the emmision region. Determination of $N_e$ and $T_e$ from the forbidden line
intensities was discussed origanally for the general case by Seaton \cite{seaton} and Seaton and Osterbrock \cite{seaton1}.
A number of such transitions have been observed in the ultraviolet
spectrum of the solar corona. Forbidden atomic emmision lines are commonly observed in quasars with an intensity often 
comparable to accompanying `allowed transition' \cite{burbidge}. Moreover, gaseous nebulae exhibit in their spectra forbidden transition lines of low excitation energy.
Many astrophysical phenomena like coronal heating, evolution of chemical composition in stellar envelopes, determination of the
chemistry in the planetary nebulae precursor's envelope are believed to be explained largely by these forbidden lines.
In laboratory tokamak plasmas and in various astronomical objects, suitably chosen these forbidden lines serve as a basis
for reliable electron density and temperature diagnostics \cite{biemont}.

Zn has been idetified in Be stars \cite{Danezis} and experimental data for absorption lines from the ground level 
are also available. Sneden et al. \cite{Sneden} had studied the abundance of Zinc (Zn) as a function of metallicity 
for a number  of metal poor stars. Zn is closely tracking ion in abundance for most stars. There are peculiar stars 
for which Zn is either scarce (if not non-existing) or over abundant \cite{smith}. An UV-atlas with line identifications 
\cite{brant} is now available for the HgMn star $\chi$-Lupi \cite{Leckrone}, which has been studied extensively. The 
signature  of singly ionized zinc (Zn II) are almost undetectable in these stars, apart from the interstellar medium (ISM). 
Zn II relatively
undepleted in the diffuse interstellar medium, and hence may be used to infer the amount of H in sightline.
The lines of Zn II have been observed in the interstellar absorption spectrum towards $\zeta$ Ophiuchi \cite{Morton}. 
The analysis data of Zn II spectrum from GHRS on the HST are done by Savage et al. 
\cite{Savage} in determining the 
depletion of Zn II. Allowed transitions and corresponding oscillator strengths of this ion have been studied
by  atomic physicsts, both experimentalists and theorists \cite{Mayo and references theirin}, but 
there has been no work towards the `forbidden transition' of Zn II till date in our best of knowledge, 
which are very important in astronomy.

Here, we have employed the relativistic coupled cluster method with single and double excitations (RCCSD) for  
determinating the `forbidden transition' amplitudes.
This is one of the most powerful highly correlated approaches due to its all order behavior for the 
correlation operator \cite{lindgren}. 
The Fock-space multi-reference  coupled cluster (FSMRCC) theory for one electron attachment process 
used here has been described elsewhere \cite{lindgren,mukherjee,Haque,Pal}.
We provide a brief review of this method. The theory for a single valence system is based on the
concept of common vacuum for both the closed shell N and open shell N$\pm$1 electron systems, which
allows us to formulate a direct method for energy differences. Also,
the holes and particles are defined with respect to the common vacuum for
both the electron systems. Model space of a (n,m) Fock-space
contains determinants with $n$ holes and $m$ particles distributed within
a set of what are termed as {\em active} orbitals. For example, in this
present article, we are dealing with (0,1) Fock-space which is a complete
model space (CMS) by construction and is given by

\begin{equation}\label{eq4}
|\Psi^{(0,1)}_\mu\rangle=\sum_i {\mbox C}_{i\mu} |\Phi_i^{(0,1)}\rangle
\end{equation}
\noindent
where ${\mbox C}_{i\mu}$'s are the coefficients of $\Psi^{(0,1)}_\mu$
and $\Phi^{(0,1)}_i$'s are the model space configurations.
The dynamical electron correlation effects are introduced through the
{\em valence-universal} wave-operator $\Omega$
\cite{lindgren,mukherjee}
\begin{equation}\label{eq5}
\Omega={\{ \exp({\tilde{S}}) }\}
\end{equation}
\noindent
where
\begin{equation}\label{eq6}
{\tilde{S}}=\sum_{k=0}^m\sum_{l=0}^n{S}^{(k,l)}
                 ={S}^{(0,0)}+{S}^{(0,1)}+ {S}^{(1,0)}+\cdots
\end{equation}
At this juncture, it is convenient to single out the core-cluster
amplitudes $S^{(0,0)}$ and call them $T$. The rest of the
cluster amplitudes will henceforth be called $S$. Since $\Omega$
is in normal order, we can rewrite Eq.(\ref{eq5}) as
\begin{equation}\label{eq7}
\Omega ={exp(T)}{\{\mbox{exp}({S}) }\}
\end{equation}
In this work, single ($T_{1}, S_{1}$) and double excitations ($T_{2}, S_{2}$) are considered for $T$ and $S$  clusters operator.
Wavefunction of the system with single valence orbital {\it v}, therefore, yields the form
\begin{equation}
|\Psi_v\rangle = \Omega_{v}|\Phi_{DF}\rangle = e^{T_{1}+T_{2}}\{1+S_{1v}+S_{2v}\}|\Phi_{DF}\rangle.
\end{equation}
Triple excitations are included in the open shell RCCSD amplitude which
correspond to the correlation to the valence orbitals, by an approximation
that is similar in spirit to CCSD(T) \cite{ccsd(t)}. The approximate
valence triple excitation amplitudes are given by

\begin{equation}
{S^{(0,1)}}_{abk}^{pqr}=\frac{{\{{\overbrace{V{T}_2}}\}_{abk}^{pqr}}+{\{{\overbrace{V{S^{(0,1)}}_2}}}\}_{abk}^{pqr}}{\varepsilon_{a}+\varepsilon_{b}+\varepsilon_{k}-\varepsilon_{p}-\varepsilon_{q}-\varepsilon_{r}},\label{eq21}
\end{equation}
where ${S^{(0,1)}}_{abk}^{pqr}$ are the amplitudes corresponding to the simultaneous
excitations of orbitals $a,b,k$ to $p,q,r$, respectively;
$\overbrace{V{T}_2}$ and $\overbrace{V{\mbox S^{(0,1)}}_2}$ are the connected
composites involving $V$ and $T$, and $V$ and $S^{(0,1)}$, respectively, where
$V$ is the two electron Coulomb integral and $\varepsilon$'s are the
orbital energies.

The the transition element due to any operator $D$ can be expressed, in the RCCSD method, as
\begin{eqnarray} 
D_{fi}& = & \frac{\langle \Psi_f|D|\Psi_i\rangle} {\sqrt{{\langle \Psi_f|\Psi_f\rangle}{\langle \Psi_i|\Psi_i\rangle}}} \nonumber \\
&=& \frac{{\langle \Phi_f|\{1+{S_f}^{\dag}\}{e^T}^{\dag}De^T\{1+S_i\}|\Phi_i\rangle}}
{\sqrt{{\langle \Phi_f|\{1+{S_f}^{\dag}\}{e^T}^{\dag}e^T\{1+S_f\}|\Phi_f\rangle}
{\langle \Phi_i|\{1+{S_i}^{\dag}\}{e^T}^{\dag}e^T\{1+S_i\}|\Phi_i\rangle}}}
\end{eqnarray}
The one-electron reduced matrix elements of  M1 and E2 operators are given by \cite{berestetski}.
\begin{eqnarray}
\langle k_f||q_m^{(M1)}||k_i \rangle = \langle \kappa_f||C_m^{(1)}||\kappa_i \rangle \frac{6}{\alpha k}
\frac{\kappa_i+\kappa_f}{2} \left[ \int dr j_1(kr)(P_{f}(r)Q_{i}(r)+Q_{f}(r)P_{i}(r))\right]
\end{eqnarray}
and
\begin{eqnarray}
\langle k_f||q_m^{(E2)}||k_i \rangle & = & \langle \kappa_f||C_m^{(2)}||\kappa_i \rangle \frac{15}{ k^2}  \nonumber \\
\times && \Big[ \int dr{j_2(kr)(P_{f}(r)P_{i}(r)+Q_{f}(r)Q_{i}(r))}+j_3(kr) (\frac{\kappa_f-\kappa_i}{3}) (P_{f}(r)Q_{i}(r)+Q_{f}(r)P_{i}(r)) \nonumber \\
&+&(P_{f}(r)Q_{i}(r)-Q_{f}(r)P_{i}(r))\Big]
\end{eqnarray}
respectively. Here, $j_i$ and $\kappa_i = \pm \left(j_{i}+\frac{1}{2}\right)$ are the total angular momentum and
relativistic angular momentum quantum numbers, respectively, of the {\it i$^{th}$} electron orbital.
The quantity $C_m^{(l)}$ is the Racah tensor and $j_l(kr)$ is the spherical Bessel function of order $l$.
$P_{ki}$ and $Q_{ki}$ are the large and small radial components of the Dirac wavefunctions.  \\
The emission transition probabilities (in sec$^{^-1}$) for the E2 and M1 channels from states {\it f} to {\it i} are given by
\begin{equation}
A^{E2}_{i,f} = \frac{1.11995\times10^{18}}{\lambda^{5}(2j_f+1)}S^{E2},
\end{equation}
\begin{equation}
A^{M1}_{i,f} = \frac{2.69735\times10^{13}}{\lambda^{3}(2j_f+1)}S^{M1},\\
\end{equation}
where $S = {|{\langle \Psi_f|D|\Psi_i\rangle}|}^2$ is the transition strength for the operator D (in a.u.) and $\lambda$ (in(\AA)) is the corresponding transition wavelength. \\

\section{Results and Discussions}

We calculate the  DF wavefunctions $|\Phi{_{DF}}\rangle$ using the Gaussian-type orbitals (GTO) as given in \cite{rajat} using the basis functions of the form \\
\begin{equation}
F^{L/S}_{i,k}(r) = C_N^{L/S}r^{k}e^{-\alpha_{i}r^2}
\end{equation}
with $k=0, 1, 2, 3,....$ for s, p, d, f ..... type orbital symmetries respectively. The radial functions `$F^{L}$' and `$F^{S}$'
represent the basis functions correspond to large and small components of the Dirac orbitals.
$ C_N^{L/S}$ are the normalization constants which depend on the exponents. The universal even tempering condition has been applied to the exponents
; i.e., for each symmetry exponents are assigned as
\begin{equation}
\alpha_{i}=\alpha_0\beta^{i-1} \hspace{1in} i=1,2,.....N
\end{equation}
where N is the number of basis functions for the specific symmetry.
In this calculation, we have used $\alpha_0=0.00831$ and $\beta=2.99$. The number of basis functions used in the present calculation is
32, 32, 30, 25, 20 for $l=$ 0, 1, 2, 3, 4 symmetries, respectively.

Number of DF orbitals for different symmetries used in the RCCSD calculations
are based on convergent criteria of core correlation energy for which it satisfies numerical completeness.
There are only 11, 10, 10, 7 and 5 active orbitals including all core electrons are considered in the CCSD(T) calculations
for $l=$ 0, 1, 2, 3, 4 symmetries, respectively. We first calculate $T$ amplitudes using the RCCSD equations of closed
shell system (Zn II) and then solve the $S$ amplitudes from the open shell equation for this single valence states of Zn II.

In Table I, we present the ionisation energy and fine structure splittings and their comparision with the NIST \cite{nist} results.
Our calculated ionisation energies are in excellent agreement with the NIST values. 
The present study uses Dirac Hartee-Fock (DHF)
orbitals to form the slater determinants to represent different atomic states and includes the effects of 
Coulomb correlation using a fully {\it ab initio} all-order many-body RCCSD method.  

The oscillator strengths for transitions between different energy states depend linearly on  energy, and
quadratically on the  matrix elements as shown in Eq. (7). This shows the necessity to calculate 
transition matrix elements appropriately.  

In table III, we have given the {\it ab initio} transition probabilities ($A_{fi}$ value)  for 
the `ground to excited' and `excited to excited' states transitions, which are also astrophysically important. 
Strong transition probabilities have been observed for many E2 transitions.

\begin{table}
\caption{Ionization Potentials(IPs) and fine structure (FS) splittings
(in cm$^{-1}$) of Zn II and their comparison with NIST values. The percentage
of differences of our calculated IP results compared to NIST results are shown in
parenthesis on the side of CCSD(T) IP results. }

\begin{tabular}{lrrrr}
\hline
             & \multicolumn{2}{c}{IP} & \multicolumn{2}{c}{FS} \\\hline 
States       &  CCSD(T)        & NIST       &  CCSD(T)  &  NIST  \\
             &                 &            &           &         \\\hline
\hline
$4s_{1/2}$   &     000.00    &    000.00  & 000.00  & 000.00     \\
$4p_{1/2}$   &     48557.23   & 48481.00   &       &  \\
$4p_{3/2}$   &  49422.75    & 49355.04    & 865.52 & 874.04\\
$5s_{1/2}$   &  88379.53    & 88437.15     &          & \\
$4d_{3/2}$   &  96781.13    & 96909.74  &     &  \\
$4d_{5/2}$   &  96830.94    & 96960.40   & 49.81 & 50.66   \\
$5p_{1/2}$   &  101329.53    & 101365.93  &       &     \\
$5p_{3/2}$   &  101623.61    & 101611.43  & 294.08  & 245.50\\
$6s_{1/2}$   &  114617.73    & 114498.02 &         &    \\
$4f_{5/2}$   &  117026.80    & 117263.40 &         &    \\
$4f_{7/2}$   &  117026.39    & 117264.00 &  0.41   & 0.60 \\
$5d_{3/2}$   &  118112.35    & 117969.32 &         &     \\
$5d_{5/2}$   &  118140.33    & 117993.61 &  27.98  & 24.29\\
$6p_{1/2}$   &  119848.71    & 119888.51   &        &     \\
$6p_{3/2}$   &  119944.98    & 119959.34 & 96.27  & 70.83\\
\hline
\end{tabular}
\label{tab:results1}
\end{table}

\begin{table}[h]
\caption{Oscillator strength corresponding to allowed transitions among different low-lying states of Zn II.}
\begin{ruledtabular}
\begin{tabular}{lrrr}
State &  Experiment & Other theories & This work   \\
\hline
$4s_{1/2}$$\rightarrow 4p_{1/2}$  &  0.249(50),$^{a}$ 0.249(29),$^{b}$ 0.256(26),$^{c}$ & 0.317,$^{a}$ 0.291,$^{b}$ 0.290,$^{e}$ & 0.2621 \\
               &0.209(18),$^{d}$ 0.306(58),$^{e}$ 0.246,$^{p}$ & 0.260,$^{f}$ 0.2526,$^{g}$ 0.238,$^{h}$   &  \\
         &  & 0.2965,$^{i}$ 0.2521,$^{j}$ 0.2,$^{k}$ & \\             
  &    &  0.2856,$^{l}$ 0.262,$^{m}$ 0.268,$^{n}$ & \\
$4s_{1/2}$$\rightarrow 4p_{3/2}$  &  0.467(93),$^{a}$ 0.48(55),$^{b}$ 0.488(49),$^{c}$&0.642,$^{a}$ 0.586,$^{b}$ 0.615,$^{e}$  & 0.535\\
            & 0.513(60),$^{q}$ 0.606(53),$^{d}$ 0.62(6)$^{r}$  & 0.515,$^{f}$ 0.5159,$^{g}$ 0.487,$^{h}$ & \\
            & 0.590(84),$^{e}$ 0.406(41),$^{s}$ 0.501,$^{p}$ & 0.6046,$^{i}$ 0.5187,$^{j}$ 0.476,$^{o}$& \\
                      & & 0.41,$^{k}$ 0.5821,$^{l}$ 0.537,$^{m}$ 0.552,$^{n}$ & \\
$4p_{1/2}$$\rightarrow 4d_{3/2}$  & 0.543(70),$^{a}$ 0.588,$^{p}$& 0.951,$^{a}$ 0.78489,$^{g}$ & 0.8218 \\
       &   &  0.789,$^{h}$ 0.8477,$^{i}$ & \\
$4p_{3/2}$$\rightarrow 4d_{3/2}$  & 0.087(11),$^{a}$  & 0.0935,$^{a}$ 0.07952,$^{g}$  & 0.0868\\
                                   &                   & 0.0799,$^{h}$ 0.08593,$^{i}$ & \\
$4p_{3/2}$$\rightarrow 4d_{5/2}$  & 0.69(14),$^{a}$ 0.684(57),$^{b}$  & 0.843,$^{a}$ 0.757,$^{b}$ 0.71483,$^{g}$  & 0.7823\\
              & 0.704(75),$^{q}$ 0.258(47),$^{d}$ 0.555,$^{p}$ & 0.718,$^{h}$ 0.7721,$^{i}$ 0.67,$^{o}$ & \\
$4d_{3/2}$$\rightarrow 4f_{5/2}$  & 0.99(20),$^{a}$ 0.868,$^{p}$ & 1.08,$^{a}$ 1.057,$^{i}$  & 1.086\\
$4d_{5/2}$$\rightarrow 4f_{7/2}$  & 1.32(26),$^{a}$  & 1.02,$^{a}$ 1.009,$^{i}$  & 1.131\\
$5p_{1/2}$$\rightarrow 5d_{3/2}$  & 0.82(0.16),$^{a}$  & 0.826,$^{a}$ 0.9428,$^{i}$  & 0.9541\\
$5p_{3/2}$$\rightarrow 5d_{3/2}$  & 0.0240(48),$^{a}$  & 0.0787,$^{a}$ 0.09629,$^{i}$  & 0.0949\\
$5p_{3/2}$$\rightarrow 5d_{5/2}$  & 0.52(10),$^{a}$  & 0.708,$^{a}$ 0.8633,$^{i}$  & 0.5706\\

\end{tabular}
\end{ruledtabular}
\label{tab:results1}
\end{table}
$^{a}$Laser-produced plasma \cite{Mayo and references theirin}\\
$^{b}$LIF \cite{blagoev} \\
$^{c}$LIF + Hollow cathode lamp \cite{bergeson} \\
$^{d}$Phase shift \cite{baumann} \\
$^{e}$Beam foil \cite{hultberg} \\
$^{f}$Relativistic many-body perturbation \cite{thorne} \\
$^{g}$MCDF + Core polarization \cite{curtis} \\
$^{h}$RHF + Core polarization \cite{migdalek} \\
$^{i}$Coulomb approximation \cite{lindgard} \\
$^{j}$Relativistic Supersymmetry Quantum Defect Theory \cite{nana} \\
$^{k}$One electron approx. with core polarisation \cite{chichkov} \\
$^{l}$WKB(sinusoidaloscillation) \cite{lagmago} \\
$^{m}$CI + Core polarization \cite{harrison} \\
$^{n}$One electron approx model potential \cite{laughlin} \\
$^{o}$HF \cite{zatsarinny} \\
$^{p}$NIST \cite{nist} \\
$^{q}$Beam foil \cite{martinson} \\
$^{r}$Beam foil \cite{andersen} \\
$^{s}$Beam foil \cite{Andersen} \\

\begin{table}[h]
\caption{ Transition wavelengths and transition probabilities of Zn II.}
\begin{tabular}{llrrr}
\hline
Terms     &     &  $\lambda_{CCSD}$(\AA) & $A_{fi}$-value E2  & $A_{fi}$-value M1\\ \hline
$4d_{3/2}$&$\rightarrow 4d_{5/2}$& 2007588.68    & 2.3078E-12   & 1.326E-06 \\
          &$\rightarrow 5d_{3/2}$& 4687.96    & 4.7206E+01     &  5.6123E-01  \\
          &$\rightarrow 5d_{5/2}$& 4681.82    & 1.4785E+01      &  3.4344E-04  \\
          &$\rightarrow 4s_{1/2}$& 1031.88    &  2.1426E+04      &   \\
          &$\rightarrow 5s_{1/2}$& 11902.49    & 2.4488      &    \\
          &$\rightarrow 6s_{1/2}$& 5606.44     & 3.2714E+01      &   \\
$4d_{5/2}$&$\rightarrow 5d_{3/2}$& 4698.93      & 2.2248E+01     &  4.0621E-04  \\
          &$\rightarrow 5d_{5/2}$& 4692.76      & 5.3873E+01     &  2.9588  \\
          &$\rightarrow 4s_{1/2}$& 1032.72      & 3.1598E+04    &    \\
          &$\rightarrow 5s_{1/2}$& 11832.34      & 3.7945     &    \\
          &$\rightarrow 6s_{1/2}$& 5612.14      & 4.9539E+01        &  \\
$5d_{3/2}$&$\rightarrow 5d_{5/2}$& 3573623.79   &  2.4516E-12 &  2.3577E-07  \\
          &$\rightarrow 4s_{1/2}$& 846.65      &4.4728E+03        &  \\
          &$\rightarrow 5s_{1/2}$& 3363.28      &6.8785E+02        &  \\
          &$\rightarrow 6s_{1/2}$& 28615.41      &5.1564E-01        &  \\
$5d_{5/2}$&$\rightarrow 4s_{1/2}$& 846.45      & 6.5596E+03     &    \\
          &$\rightarrow 5s_{1/2}$& 3360.12      & 1.0266E+03    &    \\
          &$\rightarrow 6s_{1/2}$& 28388.95     & 8.0749E-01    &    \\
$4p_{1/2}$&$\rightarrow 4p_{3/2}$& 115537.36  & 1.3904E-06     & 5.8297E-03 \\
          &$\rightarrow 5p_{1/2}$& 1894.93   &      &   3.6309  \\
          &$\rightarrow 5p_{3/2}$& 1884.43    & 5.6262E+02   &  9.8766E-02  \\
          &$\rightarrow 6p_{1/2}$& 1402.69     &       &  1.6186  \\
          &$\rightarrow 6p_{3/2}$& 1400.80     & 1.5624E+02   &  5.4193E-02  \\
          &$\rightarrow 4f_{5/2}$& 1460.51     & 3.3301E+03      &    \\
$4p_{3/2}$&$\rightarrow 5p_{1/2}$& 1926.53      & 1.1498E+03  &  2.8073E-01  \\
          &$\rightarrow 5p_{3/2}$& 1915.67      & 4.8138E+02    &  6.4942E+01  \\
          &$\rightarrow 6p_{1/2}$& 1419.93      & 2.9788E+02   &  9.9683E-02  \\
          &$\rightarrow 6p_{3/2}$& 1417.99      & 1.1143E+02   &  3.1170E+01  \\
          &$\rightarrow 4f_{5/2}$& 1479.20      & 9.6132E+02    &    \\
          &$\rightarrow 4f_{7/2}$& 1479.21      & 4.3255E+03    &    \\
$5p_{1/2}$&$\rightarrow 5p_{3/2}$& 340039.01   & 1.8059E-07   &  2.2860E-04  \\
          &$\rightarrow 6p_{1/2}$& 5399.81      &      &  1.1853E-01  \\
          &$\rightarrow 6p_{3/2}$& 5371.88      & 5.6369E+01  &  2.3181E-03  \\
          &$\rightarrow 4f_{5/2}$& 6370.53      & 9.7419E+01     &    \\

\hline
\label{tab:front3}
\end{tabular}
\end{table}

\begin{table}[h]
\caption{Continuation from table III}
\begin{tabular}{llrrr}
\hline
Terms     &     &  $\lambda_{CCSD}$(\AA) & $A_{fi}$-value E2  & $A_{fi}$-value M1\\ \hline
$5p_{3/2}$&$\rightarrow 6p_{1/2}$& 5486.96      & 1.1319E+02  &  1.5547E-02  \\
          &$\rightarrow 6p_{3/2}$& 5448.11      & 4.6981E+01     &  2.2892  \\
          &$\rightarrow 4f_{5/2}$& 6492.16      & 2.5774E+01     &    \\
          &$\rightarrow 4f_{7/2}$& 6492.33      & 1.1596E+02     &    \\
$6p_{1/2}$&$\rightarrow 6p_{3/2}$&  1038663.19  & 7.3181E-09    &8.0184E-06\\
          &$\rightarrow 4f_{5/2}$& 35437.11      & 2.8051E-02     &    \\
$6p_{3/2}$&$\rightarrow 4f_{5/2}$&  34267.93  & 9.3366E-03       &     \\
          &$\rightarrow 4f_{7/2}$& 34263.15      & 4.2045E-02     &    \\
$4s_{1/2}$&$\rightarrow 5s_{1/2}$& 1131.48     &       & 1.2592E+02\\
          &$\rightarrow 6s_{1/2}$& 872.46     &       & 4.9760E+01\\
$5s_{1/2}$&$\rightarrow 6s_{1/2}$& 3811.23     &       & 2.6961\\

\hline
\label{tab:front3}
\end{tabular}
\end{table}

\section{Conclusion}
The continuing developments in astrophysical and astronomical observations demand accurate theoretical transitions data to determine stellar chemical composition. Highly correlated relativistic coupled cluster theory has been employed to study the oscillator strengths 
of the astrophysically 
important forbidden transitions, Magnetic dipole and electric quadrupole
transition amplitudes among the bound states of Zn II are important for
astronomical observations and plasma researches.  To our knowledge these are the first calculations of oscillator 
strengths for the forbidden transitions of Zn II. All the transitions are in the ultraviolet, visible 
or near infrared regions. This work will 
motivate astronomers to observe these lines of Zn II to predict the abundances of these species in 
astronomical bodies and experimentalists to verify our results.

\section{Acknowledgment}
We are thankful to  Prof. B. P. Das, Indian institute of Astrophysics, Bangalore and Prof. Debashis Mukherjee, Indian Association of Cultivation for Science, Kolkata for helpful discussions.

\end{document}